\documentclass[letterpaper]{article} 
\usepackage{aaai23}  
\usepackage{times}  
\usepackage{helvet}  
\usepackage{courier}  
\usepackage[hyphens]{url}  
\usepackage{graphicx} 
\usepackage{natbib}  
\usepackage{caption} 
\usepackage{algorithm}
\usepackage{algorithmic}
\usepackage[english]{babel}
\usepackage{booktabs}
\usepackage{enumitem}
\usepackage{threeparttable}
\usepackage{multirow}
\usepackage{textcomp}
\usepackage{caption}
\usepackage{booktabs}
\usepackage{bm}
\usepackage{algorithm}
\usepackage{algorithmic}
\usepackage{makecell}
\usepackage{url}
\usepackage{amsmath}
\usepackage{balance}
\usepackage{graphics}
\usepackage{array}
\usepackage{flushend}
\usepackage[english]{babel}
\usepackage{latexsym}
\usepackage{bbding}
\usepackage{tabularx}
\usepackage[labelformat=simple]{subcaption}

\usepackage{newfloat}
\usepackage{listings}
\DeclareCaptionStyle{ruled}{labelfont=normalfont,labelsep=colon,strut=off} 
\floatstyle{ruled}
\newfloat{listing}{tb}{lst}{}
\floatname{listing}{Listing}
\nocopyright
\title{Ultron: An Ultimate Retriever on Corpus with a Model-based Indexer}
\author {
    Yujia Zhou,\textsuperscript{\rm 1}
    Jing Yao, \textsuperscript{\rm 3}
    Zhicheng Dou, \textsuperscript{\rm 1}
    Ledell Wu, \textsuperscript{\rm 2}
    Peitian Zhang, \textsuperscript{\rm 1}
    Ji-Rong Wen \textsuperscript{\rm 1}
}
\affiliations {
    \textsuperscript{\rm 1} Gaoling School of Artificial Intelligence, Renmin University of China, Beijing, China\\
    \textsuperscript{\rm 2} Beijing Academy of Artificial Intelligence, Beijing, China\\
    \textsuperscript{\rm 3} Microsoft Research Asia, Beijing, China\\
    zhouyujia@ruc.edu.cn, jingyao@microsoft.com, dou@ruc.edu.cn
}

\begin{document}

\maketitle

\begin{abstract}
Document retrieval has been extensively studied within the \textit{index-retrieve} framework for decades, which has withstood the test of time. Unfortunately, such a pipelined framework limits the optimization of the final retrieval quality, because indexing and retrieving are separated stages that can not be jointly optimized in an end-to-end manner. In order to unify these two stages, we explore a model-based indexer for document retrieval. Concretely, we propose Ultron, which encodes the knowledge of all documents into the model and aims to directly retrieve relevant documents end-to-end. For the model-based indexer, how to represent docids and how to train the model are two main issues to be explored. Existing solutions suffer from semantically deficient docids and limited supervised data. To tackle these two problems, first, we devise two types of docids that are richer in semantics and easier for model inference. In addition, we propose a three-stage training workflow to capture more knowledge contained in the corpus and associations between queries and docids. Experiments on two public datasets demonstrate the superiority of Ultron over advanced baselines for document retrieval.
%
\end{abstract}

\section{Introduction}
\label{sec:intro}
Search engines are helpful to meet users' daily information needs. Given an issued query, search engines first retrieve candidate documents from a large document collection and then re-rank these candidate documents to return a ranking list~\cite{Robertson2009BM25,Zhan2020RepBERT,Mitra2018DenseSurvey}. In this process, the performance of the retrieval stage is critical to the final search quality. For document retrieval, the inverted index has served as the backbone of term-based sparse retrieval over the past decades. With an inverted index built on the corpus, sparse retrieval methods such as BM25~\cite{Robertson2009BM25} can measure term frequency and other lexical features to retrieve relevant documents, which encounters the challenge of vocabulary mismatch. In recent years, many dense retrieval methods are proposed to tackle this problem~\cite{Gao2021Sparse_Dense,Zhan2020RepBERT}. They first encode the semantic information contained in queries and documents into dense vectors, and then retrieve relevant documents over the vectorized index.

Previous methods, including both sparse and dense retrieval models, are extensively studied within the \textit{index-retrieve} framework that has been proven to be valuable for document retrieval. Unfortunately, such a pipeline-based framework requires a large pre-computed index built on the whole corpus to support subsequent document retrieval. This not only results in huge memory overhead, but also limits optimizing the separated indexing and retrieval stages in an end-to-end way. To address these limitations, several studies~\cite{Tay2022DSI, Bevilacqua2022SEAL, Zhou2022DynamicRetriever} have preliminarily tried to build an end-to-end retrieval model that directly returns relevant documents without separated indexes. Semantic information of documents in the corpus is encoded into a large model, which can be regarded as a differentiable indexer and can be optimized end-to-end. Inspired by this framework, we intend to delve deeper into the challenges and explore better usability of this model-centric paradigm for document retrieval tasks.

Within such a model-centric retrieval framework, there are two main technical difficulties: (1) how to represent docids so that the model can learn the semantics of documents and retrieve the correct docids more easily; (2) how to train the model so as to capture the semantic knowledge of each docid and to learn the mapping relations from queries to relevant docids. Some early studies, such as DSI~\cite{Tay2022DSI}, tried a variety of inspiring document identifiers, such as atomic docids, semantic cluster docids, etc. However, these heuristic docids lack sufficient document semantic knowledge, which plays a critical role in document retrieval. In addition, existing models are only trained with limited supervised data. This limits the model to learn sufficient knowledge over each docid for retrieval and might lead to model over-fitting. In this paper, we follow the model-centric paradigm for document retrieval and attempt to deal with the above two main challenges. 

To enrich the semantic information of docids, we represent each docid as a sequence of tokens that can summarize the underlying semantics of the document. Motivated by traditional term-based index and vectorized index, we devise two types of docids corresponding to the sparse and dense retrieval respectively. The first is \textbf{Keyword-based identifiers}, which identify documents using a sequence of keywords. In this paper, we use the URL and title of webpages, which are natural keywords that somewhat guarantee both the uniqueness and semantics of the identifier. The second is \textbf{Semantic-based identifiers}, which represent a document with a series of latent semantic tokens. Inspired by the effectiveness of product quantization (PQ) technology \cite{Ge2014OPQ, Jegou2011PQ, Zhan2021JPQ} in retrieval, we use the PQ code of a document as a type of semantic-based identifier.

For model training, the training target of retrieval is to learn the mapping relations from queries to relevant docids. To cope with the problem of limited click data, we devise a three-stage training workflow, which adds two stages of pre-training to enrich the knowledge of docids stored in the model-based indexer. (1) \textbf{General Pre-training}. This stage aims to capture the document's semantic information by mapping passages and key terms of the document to the corresponding docid. 
(2) \textbf{Search-oriented Pre-training}. To enhance the performance on search tasks, the model needs to be more capable of mapping short query-like texts to corresponding docids. To get sufficient data, we generate a lot of pseduo queries for model pre-training, thereby adapting the model to search scenarios. (3) \textbf{Supervised Fine-tuning}. The final stage is to finetune the model on supervised relevance data, so as to learn more robust associations between queries and relevant docids.

Specifically, we propose the model Ultron: an \textbf{ult}imate \textbf{r}etriever \textbf{on} corpus with a model-based indexer, which is built on a generative language model with transformer-based encoder-decoder. It regards document retrieval as a sequence-to-sequence task from an issued query to docids. In the pre-processing stage, we generate two types of semantic identifiers for each document in the corpus. During training, we apply the three-stage training to encode the semantic knowledge of documents into the model parameters. At the inference time, given a query, Ultron generates a docid ranking list directly based upon the generation probabilities via constrained beam search. Experimental results on MS MARCO and NQ datasets indicate that our Ultron model not only outperforms advanced baselines of sparse or dense retrieval, but also achieves significant improvements over existing end-to-end retrieval methods.

The contributions of this work can be summarized as: (1) Along with the blueprint of model-based IR~\cite{Metzler2021forum}, we propose Ultron, a sequence-to-sequence model that directly generates docids for a query. It demonstrates the feasibility of model-based indexer on the document retrieval task. (2) To enhance the semantics of docid, we devise keyword-based and semantic-based docids, which show richer semantic attributes than existing methods. (3) To alleviate the problem of limited supervised data, we propose a three-stage training workflow to encode more knowledge over docids into the model for better document retrieval.

\section{Related Work}

\subsubsection{Sparse Retrieval with Inverted Indexes.}
Thanks to efficiency and effectiveness, sparse retrieval with an inverted index is widely used in practice. BM25~\cite{Robertson2009BM25} employs the tf-idf signal to measure term weights and compute relevance scores. Graph-based approaches~\cite{Blanco2012Graph,Rousseau2013Graph} build document graphs and follow the idea of PageRank to calculate term weights. With representation learning~\cite{Mikolov2013Word2vec,Pennington2014Glove}, one line of research~\cite{Zheng2015WordEmbedding,Guo2016WordEmbedding,Dehghani2017WordEmbedding} automatically learns term weights from word embeddings rich in semantic and co-occurrence information. DeepCT~\cite{Dai2019DeepCT} and HDCT~\cite{Dai2020HDCT} leverage contextualized text representation to predict term importance. A challenge of sparse retrieval is the mismatch between query and document words. One kind of methods~\cite{Nogueira2019Doc2Query,Nogueira2019DocT5Query} expands possible terms for queries or documents. A second approach utilizes word vectors to measure soft similarity between terms. 

\subsubsection{Dense Retrieval with Vectorized Indexes.}
These methods rely on deep learning to capture the semantic relevance between queries and documents, going beyond lexical overlap to relieve the mismatch problem~\cite{Gao2021Sparse_Dense}. They first embed queries and documents into vectors. Then, relevant documents are retrieved based on the vector similarity~\cite{Zhan2020RepBERT,Ni2021T5Dual,Karpukhin2020dpr}, where ANNS and PQ~\cite{Jegou2011ANN} are used to improve the efficiency. With advanced PLMs~\cite{Devlin2019BERT,Raffel2020T5}, higher-quality representations are obtained, leading to better results. Besides, negative sampling strategies have been proposed for better optimization~\cite{Zhan2021HardNeg,Xiong2021HargNeg,Gao2021Sparse_Dense,Guu2020HardNeg}. Considering the retrieval performance may be bounded by the product between single vectors, lightweight interaction layers are introduced to capture fine-grained matching, such as the multi-vector model~\cite{Luan2021MultiEncoder} and ColBERT~\cite{Khattab2020ColBERT}.

In this paper, we replace the traditional indexes with a model-based indexer, which can be optimized end-to-end during the model training.

\begin{figure*}
    \centering
    \vspace{-0.2cm}
    \includegraphics[width= 0.8\linewidth]{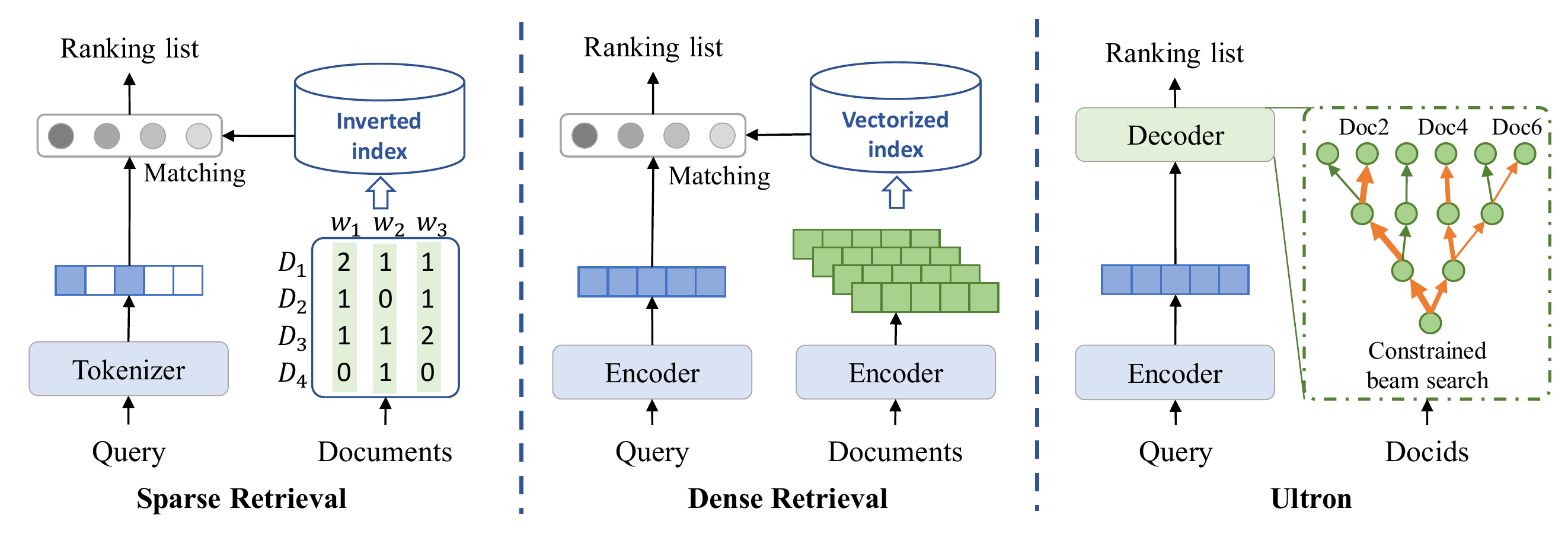}
    \caption{The comparison between Sparse Retrieval, Dense Retrieval, and Ultron. Traditional retrieval methods follow the index-retrieve framework, while the Ultron only uses an end-to-end model to accomplish document retrieval. The docids are generated from the query with a seq-to-seq model, and the ranking list is obtained through the constrained beam search.}
    \label{fig:backbone}
\end{figure*}

\subsubsection{End-to-End Retrieval.}
Different from pipelined retrieval methods relying on a separated index, end-to-end models without explicit indexes have been applied to retrieval tasks. GENRE~\cite{Cao2021genre} retrieves entities by generating their names through a seq-to-seq model. A blueprint for model-based IR is proposed in~\cite{Metzler2021forum}, which aims to embed the knowledge of all documents into a model. Under this framework, some studies explored the document retrieval task. \citet{Tay2022DSI} devise DSI to directly output docids for document retrieval. \citet{Bevilacqua2022SEAL} regard all n-grams in the passage as identifiers and generate relevant words to retrieve documents. \citet{Chen2022gene} propose GERE to retrieve evidence by returning sentence identifiers. We observe that existing works suffer from semantically deficient docids and limited supervised data. In this paper, we continue to explore a more effective retriever within this model-centric paradigm.


\section{Methodology}
In this paper, we explore the model-centric paradigm for document retrieval, and attempt to deal with two main technical difficulties: (1) how to represent docids so that the model can easily learn the semantics of them for decoding; (2) how to train the model so as to capture knowledge of docids for retrieval. To this end, we propose Ultron, an end-to-end retriever with two elaborate docids and a three-stage training strategy. We will introduce Ultron in the next parts.


\subsection{Backbone of the Model}
\label{sec:model}
Inspired by advanced generative PLMs~\cite{Brown2020GPT3,Raffel2020T5}, we complete the document retrieval task in a generative manner through a seq-to-seq model. As shown in Figure~\ref{fig:backbone}, Ultron is implemented under an encoder-decoder framework, which encodes the input query and decodes relevant docids through constrained beam search to generate a ranking list. Compared with the traditional sparse and dense retrieval, Ultron turns the matching task into a generation task, which breaks away from the indexing-matching workflow. Such a shift eliminates traditional indexes and enables the model to be optimized end-to-end.

\textbf{Sequence-to-sequence Model}.
Since the seq-to-seq structure works well for many generation tasks, we take advantage of the T5~\cite{Raffel2020T5} pre-trained language model as our backbone, which is a Transformer-based~\cite{Vaswani2017Transformer} encoder-decoder structure. For Ultron, we define the basic task as a ``text-to-docid'' format, which means the model takes in some text and generates relevant docids (represented as a sequence of tokens). Based on the query $q$, the model tries to predict relevant docids with the highest auto-regressive scores, denoted as:
\begin{equation}
    \text{score}(d|q) = p_\theta(y|q) = \prod^N_{i=1} p_\theta(y_i|y_{<i}, q),
\end{equation}
where $y$ is the string docid of the document $d$ with $N$ tokens, and $\theta$ is model parameters. Unlike traditional seq-to-seq tasks, free-form generation might result in an output string that does not hit any valid docids. We resolve this problem in the following part.

\textbf{Constrained Beam Search}.
Beam search is a typical decoding algorithm that improves greedy search. However, since we need to ensure that the generated docids exist in the corpus, vanilla beam search cannot meet our needs. Motivated by \cite{Cao2021genre}, we apply constrained beam search to guide the decoder to search in a limited token space at each step, so as to generate valid docids. Specifically, we define the constraints based on a prefix tree built on all docid strings. Each node in the tree denotes a specific prefix sequence, and its child nodes constitute all valid subsequent tokens. By decoding along the prefix tree, the model can ensure that the generated docids exist in the corpus. Finally, according to the auto-regressive scores during beam search, the model generates the top-k docids as the ranking result.

\begin{figure}
    \centering
    \includegraphics[width=1.0\linewidth]{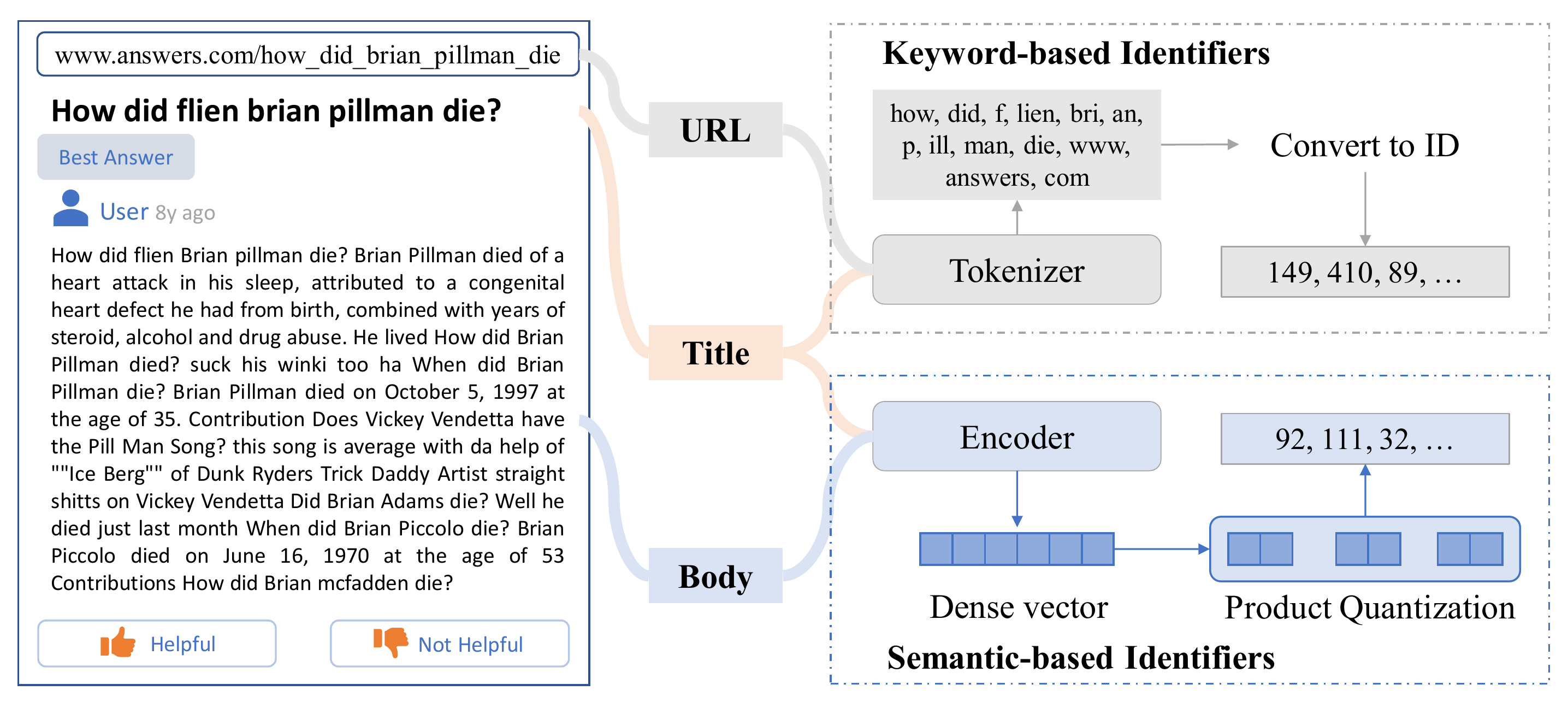}
    \caption{Two types of semantic document identifiers.}
    \label{fig:docid}
\end{figure}

\begin{figure*}
    \centering
    \vspace{-0.2cm}
    \includegraphics[width=0.8\linewidth]{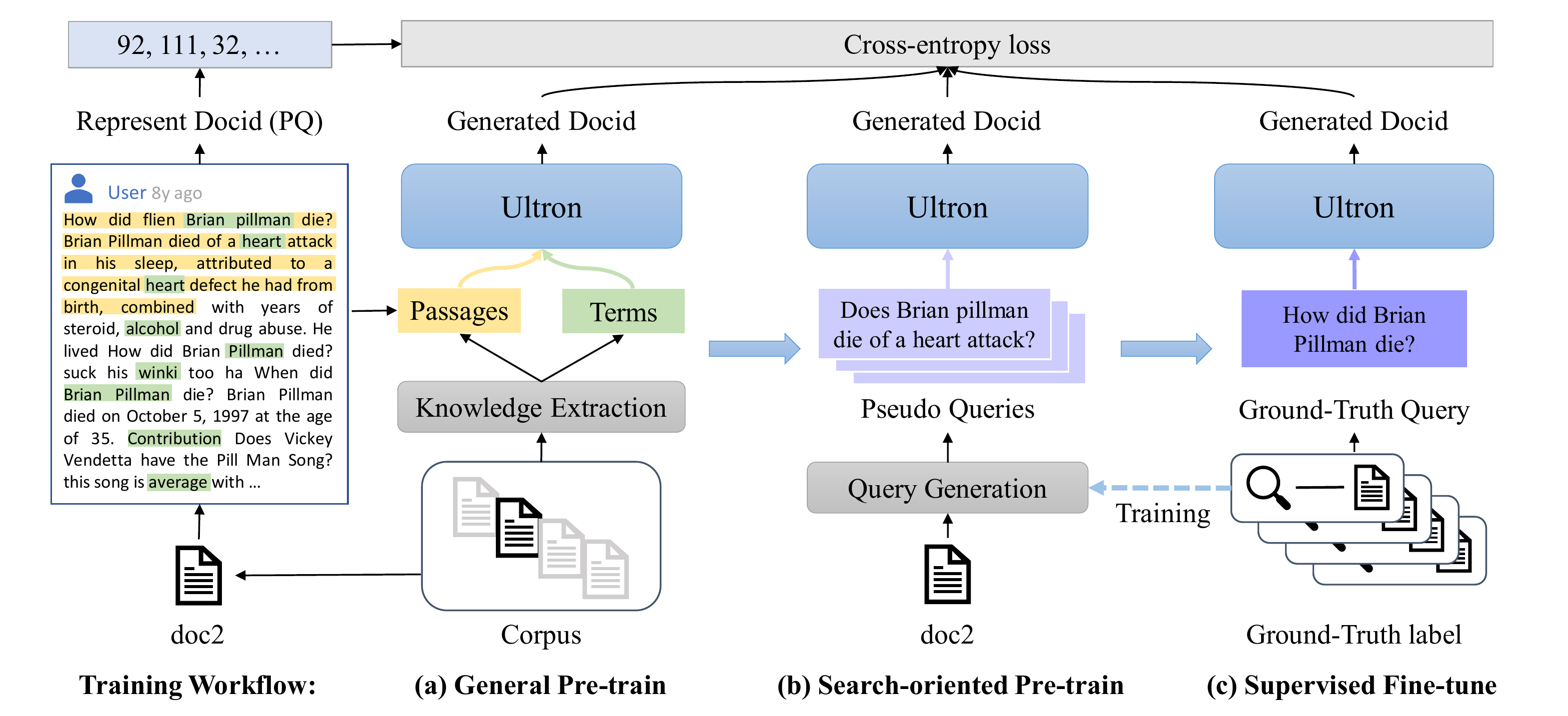}
    \caption{The three-stage training workflow of the Ultron model.}
    \label{fig:training}
\end{figure*}

\subsection{Design of Document Identifiers}
\label{sec:docid}
A natural attribute of document identifiers (docid) is to distinguish different documents. Intuitively, previous works have tried to identify documents with a random integer, called atomic identifiers~\cite{Tay2022DSI,Zhou2022DynamicRetriever}. However, they lead to gigantic embedding parameters and lack semantics. To alleviate this problem, we represent each docid as a sequence of shareable tokens satisfying two characteristics: uniquely referring to a document and reflecting the semantic information of the document. Following the ideas of sparse retrieval and dense retrieval, as shown in Figure~\ref{fig:docid}, we attempt to represent docids from two perspectives: keyword-based identifiers and semantic-based identifiers.

\textbf{Keyword-based Identifiers}.
Using keywords to represent the document content is a classic way of sparse retrieval. Motivated by this, we propose to uniquely identify a document with some meaningful keywords. Intuitively, the URL of a document contains certain semantic information and can uniquely correspond to this document. This inspires us to use URLs as docids and generate relevant document URLs directly for a given query. To facilitate model prediction, we arrange each part of the URL (split by `/') in reverse order, so as to predict the semantic-rich part first, and then predict the domain name. However, not all URLs provide sufficient semantic information. To deal with such situations, we incorporate the document title to represent docids, defined as:
\begin{align*}
\begin{split}
\text{docid}_{\text{URL}}= \left \{
\begin{array}{ll}
    \text{reversed URL}, & \text{if title length} \leq L,\\
    \text{title} + \text{domain}, & \text{otherwise.}
\end{array}
\right.
\end{split}
\end{align*}
Here $L$ is set to 2 in our experiments. Finally, we can get a sequence of tokens by T5 tokenizer to represent the docid.

\textbf{Semantic-based Identifiers}.
Dense retrieval maps documents into latent semantic space as dense vectors. In extreme cases, each dense vector can be used as a unique identifier to distinguish documents. However, the space of dense vectors is too large to decode. This promotes us to look for a way to preserve dense vector semantics in a smaller space. As a classic vector compression method, Product Quantization \cite{Jegou2011PQ, Ge2014OPQ, Zhan2021JPQ} just meets our needs for designing docids. For all D-dimensional vectors, it first divides the D-dimensional space into $m$ groups, and then performs K-means clustering on each group to obtain $k$ cluster centers. Finally, each vector can be represented as a set of $m$ cluster ids. Thus, for the document $d$, its semantic-based identifier can be defined as:
\begin{equation}
    \text{docid}_{\text{PQ}} = \text{PQ}\left(\text{Encoder}\left(d\right)\right),
\end{equation}
where $\text{Encoder}(\cdot)$ is implemented by a pre-trained T5 encoder~\cite{Ni2021T5Dual}. For cluster ids of all groups, we regard them as $m \times k$ new tokens and add them into the vocabulary.

\subsection{Training Workflow}
The training of Ultron can be regarded as the process of building a model-based indexer over the corpus. Through this process, we expect the model to encode rich semantics over docids and learn the mapping relations from queries to relevant docids. However, insufficient supervised click data makes it hard for the model to learn associations between queries and docids. This motivates us to construct more self-supervised query-docid pairs to adapt the model to search scenarios. As shown in Figure~\ref{fig:training}, the training of Ultron is divided into three stages: general pre-training, search-oriented pre-training and supervised fine-tuning. The details of the three stages are introduced in the following parts.

\textbf{General Pre-training}.
The semantic information contained in the document is the basic knowledge of the docid, which is generally useful in IR tasks. To learn such knowledge, we conduct general pre-training by extracting self-supervised signals from the corpus. Specifically, term sequences are extracted from the document content to construct the mapping relations from text to docid. We conduct two simple but effective strategies to achieve this.

First, inspired by previous studies that use passage-level evidence for document ranking~\cite{Callan1994passage}, we segment the document into passages by fixed-size windows, and construct \textit{passage-to-docid} samples for model training. Formally, given a document containing $n$ terms, i.e. $\{t_1, t_2, ..., t_n\}$, we can extract a batch of training pairs with length of size $s$, such as:
\begin{equation}
    \text{passage}: \{t_i, t_{i+1}, ..., t_{i+s}\} \longrightarrow \text{docid},
\end{equation}
where $i$ is any starting position and is set at intervals of $s$.

Second, the importance of each word in a document for its semantic representation is different~\cite{Robertson2009BM25}. Thus, we can select some important words based on tf-idf weights as a term set to reflect the key semantics of the document. We have:
\begin{equation}
    \text{terms}: \{t_i, ...,t_j, ..., t_k\} \longrightarrow \text{docid},
\end{equation}
where $t_i, t_j, t_k$ are terms selected from the document. 


\textbf{Search-oriented Pre-training}. 
After general pre-training, the model already understands the basic semantics of each docid, but we observe that this is not enough for the model to perform well on document retrieval. In other words, in addition to the semantic knowledge of documents, the model needs to further learn the associations between queries and docids. To adapt the model to search scenarios, we further conduct search-oriented pre-training, which generates pseudo queries based on the corpus and learns the mapping relations from queries to docids.

Following DocT5Query~\cite{Nogueira2019DocT5Query}, we first train a query generator over supervised click data based on a T5 model. Then, for a document, we take the first passage as the input, and the query generator outputs $k$ predicted queries with beam search, i.e. $Q=\{q_1,...,q_k\}$. Finally, by training over \textit{pseudo query-to-docid} samples, our model learns the mapping relations from query-like strings to docids, implementing the adaptation from general tasks to retrieval tasks. Formally, the training pairs are formed as:
\begin{equation}
    \text{pseudo query}: q_i \longrightarrow \text{docid}, i \in \{1,...,k\}.
\end{equation}

\textbf{Supervised Fine-tuning}.
After general and search-oriented pre-training, our model has certain knowledge and retrieval ability over docids. In order to adapt the model to specific data distribution of the downstream dataset, we further use supervised data to finetune the model. Specifically, the supervised data is query-relevant docid pairs. By training the model on \textit{query-to-docid} samples, it is aware of rich knowledge for document retrieval.

Since the above training tasks are unified in a ``text-to-docid'' format, we complete the three-stage training of Ultron based on the standard seq-to-seq objective, i.e., maximizing the target sequence likelihood with teacher forcing. Concretely, for the input sequence $q$, the generation objective can be formalized as:
\begin{equation}
    \mathcal{L}=\mathop{\arg\max}\limits_{\theta} \sum_{i} \log p_\theta(y_i|y_{<i}, q),
\end{equation}
where $p_\theta(y_i|y_{<i}, q)$ is the generation probability of token $y_i$ based on the given input. The parameters are optimized by the cross-entropy loss and the AdamW optimizer.

\section{Experimental Settings}

\subsection{Datasets}\label{subset:data}
We experiment with the following two public datasets for document retrieval.

\textbf{MS MARCO Document Ranking}\footnote{\url{https://microsoft.github.io/msmarco/Datasets}}~\cite{Nguyen2016MSMARCO} is a large collection of queries and web-pages. Following the previous work~\cite{Zhou2022DynamicRetriever}, we construct a document subset sampled from the labeled documents, and use their corresponding queries for training. There are about 320k documents and 360k query-document pairs in this subset.

\textbf{Natural Question 320K}\footnote{\url{https://ai.google.com/research/NaturalQuestions/download}}~\cite{Kwiatkowski2019NQ} is a public dataset. Each piece of data contains a real question and a Wikipedia article to answer this question. We use URLs to deduplicate documents in the corpus and the collection remains about 231k articles with 307k training pairs and 7.8k test queries. More details of data processing are introduced in supplementary materials.

\begin{table*}[!ht]
    \centering
    \small
    \caption{Overall results. ``(-)'' and ``($\uparrow$)'' indicates whether the params will increase as the corpus size increases. The best results are shown in \textbf{bold} and the best results of (-) models are \underline{underlined}. ``$\ddagger$'' and ``$\dagger$'' denotes the result is significantly better than all baselines and (-) baselines in t-test with $p \textless 0.05$.}
    \begin{tabular}{p{0.14\linewidth}|p{0.075\linewidth}<{\centering}|p{0.065\linewidth}<{\centering}p{0.065\linewidth}<{\centering}p{0.065\linewidth}<{\centering}p{0.075\linewidth}<{\centering}|p{0.065\linewidth}<{\centering}p{0.065\linewidth}<{\centering}p{0.065\linewidth}<{\centering}p{0.075\linewidth}<{\centering}}
    \toprule
        \multirow{2}[2]{*}{Model} & \multirow{2}[2]{*}{Params} & \multicolumn{4}{c|}{MS MARCO} & \multicolumn{4}{c}{Natural Questions (NQ)} \\
    \cmidrule(lr){3-10}
        & & R@1 & R@5 & R@10 & MRR@10 & R@1 & R@5 & R@10 & MRR@10 \\
    \midrule
        \multicolumn{10}{l}{Sparse Retrieval} \\ 
    \midrule
        BM25 & - & 0.1894 & 0.4282 & 0.5507 & 0.2924 & 0.1406 & 0.3691 & 0.4793 & 0.2360 \\
        DocT5Query & - & 0.2327 & 0.4938 & 0.6361 & 0.3481 & 0.1907 & 0.4388 & 0.5583 & 0.2955 \\
    \midrule
        \multicolumn{10}{l}{Dense Retrieval} \\ 
    \midrule
        RepBERT &   220M (-) & 0.2525 & 0.5841 & 0.6918 & 0.3848 & 0.2257 & 0.5220 & 0.6565 & 0.3513\\
        Sentence-T5 &   220M (-) & 0.2727 & 0.5891 & 0.7215 & 0.4069 & 0.2251 & 0.5200 & 0.6512 & 0.3495 \\
        DPR &   220M (-) & 0.2908 & 0.6275 & 0.7313 & 0.4341 & 0.2278 & 0.5344 & \underline{0.6858} & 0.3592\\
    \midrule
        \multicolumn{10}{l}{End-to-end Retrieval} \\ 
    \midrule
        DSI-Semantic &  250M (-) & 0.2574 & 0.4358 & 0.5384 & 0.3392 & 0.1323 & 0.3701 & 0.4828 & 0.2377\\
        DSI-Atomic & 495M ($\uparrow$) & 0.3247 & 0.6301 & 0.6992 & 0.4429 & 0.2023 & 0.4872 & 0.6146 & 0.3216 \\
        DynamicRetriever & 495M ($\uparrow$) & 0.2904 & 0.6422 & 0.7315 & 0.4253 & 0.2263 & 0.5353 & 0.6876 & 0.3608 \\
    \midrule
        Ultron-URL &   248M (-) & 0.2957$^\dagger$ & 0.5643 & 0.6782 & 0.4002 & \underline{\textbf{0.3378}}$^\ddagger$ & \underline{0.5420}$^\ddagger$ & 0.6705 & \underline{\textbf{0.4251}}$^\ddagger$\\
        Ultron-PQ &   257M (-) & \underline{0.3155}$^\dagger$ & \underline{0.6398}$^\dagger$ & \underline{0.7314} & \underline{0.4535}$^\ddagger$ & 0.2564$^\ddagger$ & 0.5309 & 0.6575 & 0.3712$^\ddagger$\\
        Ultron-Atomic & 495M ($\uparrow$) & \textbf{0.3281}$^\ddagger$ & \textbf{0.6490}$^\ddagger$ & \textbf{0.7413}$^\ddagger$ & \textbf{0.4686}$^\ddagger$ & 0.2543$^\ddagger$ & \textbf{0.5482}$^\ddagger$ & \textbf{0.6953}$^\ddagger$ & 0.3859$^\ddagger$\\
    \bottomrule
    \end{tabular}
    \label{tab:overall}
\end{table*}

\subsection{Baselines}
For comparison, we select three classes of baseline models.

(1) \textbf{Sparse Retrieval}.
These methods score candidate documents based on the weight of appearing query terms. \textit{BM25}~\cite{Robertson2009BM25} uses the tf-idf feature to measure term weights, implemented under Lucene. \textit{DocT5Query} \cite{Nogueira2019DocT5Query} expands a document with possible queries predicted by a finetuned T5~\cite{Raffel2020T5} with this document as the input.

(2) \textbf{Dense Retrieval}.
We mainly focus on the dual encoder framework with large-scale PLMs as the backbone. Two implementations with different underlying encoders, i.e. \textit{RepBERT}~\cite{Zhan2020RepBERT} and \textit{Sentence-T5}~\cite{Ni2021T5Dual} are considered. \textit{DPR}~\cite{Karpukhin2020dpr} samples hard negatives for optimization. 

(3) \textbf{End-to-end Retrieval}.
DSI~\cite{Tay2022DSI} uses a text-to-text model to map queries to relevant docids. We reproduce \textit{DSI-Atomic} and \textit{DSI-Semantic} for comparison. \textit{DynamicRetriver}~\cite{Zhou2022DynamicRetriever} deploys a Docid decoder with a trainable vector for each document. We reproduce the OverDense variant. \textit{Ultron-Atomic}, \textit{Ultron-URL} and \textit{Ultron-PQ} are three variants of Ultron with atomic docids, URL docids and PQ docids respectively.

We evaluate the performance of all models on the common evaluation metrics for document retrieval Recall (R@1/5/10), and Mean Reciprocal Rank (MRR@10).

\subsection{Implementation Details}
In our experiments, BERT corresponds to the pre-trained bert-base-uncased one and T5 is t5-base, both from huggingface transformers. 
As for RepBERT, Sentence-T5 and DPR, the max length of input sequences is 512, and the batch size is 48. 
We follow the settings of DynamicRetriver~\cite{Zhou2022DynamicRetriever} and DSI~\cite{Tay2022DSI} in their original papers. For Ultron, the max length of URL docids is 100, the hyper-parameter of PQ is $m=24, k=256$, and the batch size is set as 128. During the three-stage training, we utilize 10 pieces of passage, 1 key term sequence, 10 pseudo queries and 1 annotated finetune query for each document, and the learning rate is 1e-3. During inference, the beam size is 10. All models are trained with the AdamW~\cite{DLoshchilov2019AdamW} optimizer. More implementation details and source code are provided in supplementary materials. 

\section{Experimental Results}\label{sec:results}
In this section, we conduct extensive experiments to answer the following research questions:
\begin{itemize}[leftmargin=*]
    \item \textbf{RQ1}: How does Ultron perform on document retrieval compared to existing static index-based methods?
    \item \textbf{RQ2}: How does each stage of the three-stage training workflow contribute to the final retrieval results?
    \item \textbf{RQ3}: Does Ultron have lower memory overhead and higher inference speed than existing retrieval methods?
    \item \textbf{RQ4}: What are the advantages of Ultron's document identifiers over the semantic cluster docids in DSI?
\end{itemize}

\subsection{Overall Performance (RQ1)}
We present the overall results in Table~\ref{tab:overall}, and draw several conclusions as follows to answer \textbf{RQ1}.

(1) In most scenarios, the end-to-end retrieval models outperform static index-based retrieval methods, with paired t-test at $p \textless 0.05$ level. On MS MARCO, Ultron-PQ exceeds DPR by 8.5\% on R@1 and 4.5\% on MRR@10. Ultron-Atomic achieves the best performance on both the MS MARCO and NQ datasets. We think the advantage of end-to-end retrievers lies in that their indexing and retrieval stages can be jointly optimized end-to-end to better fit the final retrieval target. Furthermore, our proposed Ultron model performs the best among all end-to-end retrievers. This is thanks to our designed docids rich in semantics and the three-stage training workflow that better captures the mapping relations from queries to docids to fit search scenarios.

(2) Comparing the three end-to-end retrievers with different semantic docids, the results of Ultron-URL and Ultron-PQ are better than that of DSI-Semantic. For the NQ dataset, Ultron-URL improves DSI-Semantic by 38.9\% on R@10. This indicates that our devised docids can embed richer semantic information to facilitate model training and decoding. On MS MARCO dataset, Ultron-URL is inferior to Ultron-PQ. Ultron-URL takes URLs as docids and tends to capture the semantic knowledge related to these keywords. Some noises in URLs, such as numbers and symbols, may affect capturing semantics. Whereas the PQ docids are generated from the representation of the entire document, thus better associated with the document knowledge.

(3) End-to-end retrievers with atomic docids (including DSI-Atomic, DynamicRetriver and Ultron-Atomic) achieve better results than those with semantic docids (Ultron-PQ/URL) on most metrics. This indicates that more parameters make it easier to distinguish different documents. In addition, Ultron-Atomic learns more information from pseudo query-docid pairs, outperforming DSI-Atomic and DynamicRetriever. However, this kind of docids could lead to gigantic parameters and memory burden, especially when the number of documents increases. The semantic docids using sharable tokens have the potential to alleviate these issues.

In summary, these results indicate that \textbf{our end-to-end retriever Ultron with semantic-richer document identifiers is promising for document retrieval tasks.}

\begin{table}[!t]
    \centering
    \small
    \caption{Ablation study of the three-stage training workflow.}
    \begin{tabular}{p{0.32\linewidth}|p{0.12\linewidth}<{\centering}p{0.11\linewidth}<{\centering}|p{0.12\linewidth}<{\centering}p{0.11\linewidth}<{\centering}}
    \toprule
        \multirow{2}[2]{*}{Model} & \multicolumn{2}{c|}{MS MARCO} & \multicolumn{2}{c}{Natural Questions} \\
    \cmidrule(lr){2-5}
        & MRR@10 & R@10 & MRR@10 & R@10 \\
    \midrule
        Ultron-URL & \textbf{0.4002} &  \textbf{0.6782} & \textbf{0.4251} & \textbf{0.6705}  \\ 
    \midrule
        ~ w/o general pretrain & 0.3856 & 0.6321 & 0.3587  & 0.6608  \\
        ~ w/o search-oriented & 0.3341 & 0.5211 & 0.3071 & 0.6147 \\
        ~ w/o finetune & 0.3477 & 0.5693 & 0.3504 & 0.6405  \\
    \midrule
        Ultron-PQ & \textbf{0.4535} & \textbf{0.7314} & \textbf{0.3712} & \textbf{0.6575} \\ 
    \midrule
        ~ w/o general pretrain & 0.4099 & 0.6968 & 0.3328 & 0.6327 \\
        ~ w/o search-oriented & 0.3445 & 0.5730 & 0.2427 & 0.5220 \\
        ~ w/o finetune & 0.4176 & 0.7023 & 0.3522 & 0.6386 \\
    \bottomrule
    \end{tabular}
    \label{tab:ablation}
\end{table}

\subsection{Study of Training Workflow (RQ2)}
In this paper, we design a three-stage training workflow for Ultron. To verify the effects of each training stage on the final results (\textbf{RQ2}), we conduct an ablation study to remove one training stage at one time and observe the impacts on document retrieval. The results are shown in Table~\ref{tab:ablation}. 

We find that the removal of each training stage will damage the results on all evaluation metrics. Concretely, removing the search-oriented pre-training causes the biggest drop in overall results. This indicates that the pseudo query-docid pairs significantly enhance the model's performance on search tasks. Meanwhile, the general pre-training also contributes to the final results. This stage is dedicated to gaining the document knowledge of each docid. After removing the supervised fine-tuning, the performance on document retrieval drops a lot. This result confirms the necessity of the only supervised stage and its consequence of learning more robust associations from queries to docids.

\begin{figure}[!t]
    \centering
    \vspace{-0.1cm}
    \includegraphics[width=0.9\linewidth]{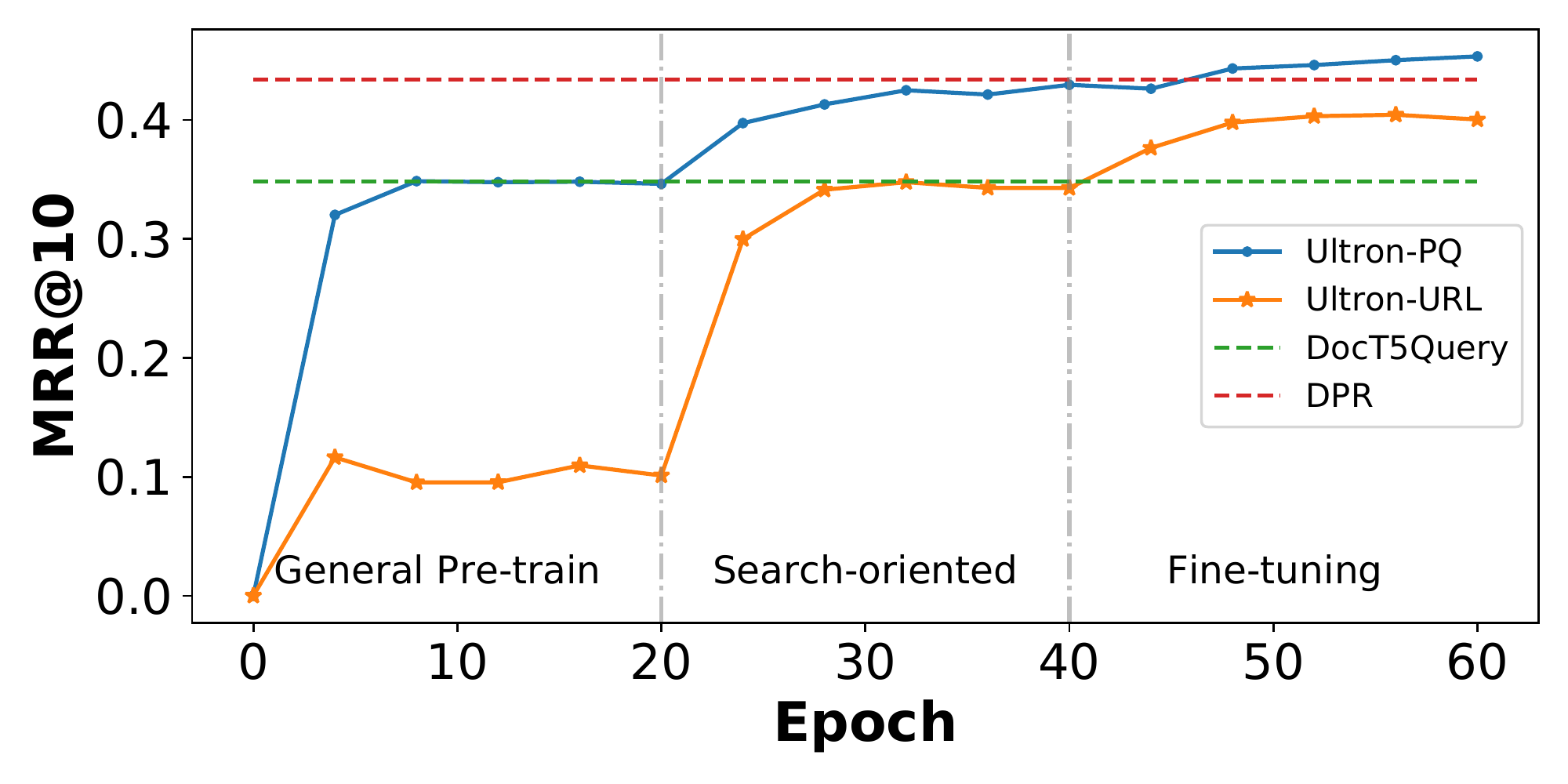}
    \caption{Training curves on MS MARCO dataset.}
    \label{fig:stage_curve}
\end{figure}

To illustrate the effects of each training stage in a finer-grained manner, we plot the training curve on MS MARCO with MRR@10 against the training epochs, shown in Figure~\ref{fig:stage_curve}. With each stage, the model gradually captures knowledge to better complete the document retrieval task. Particularly, the general pre-training contributes more to Ultron-PQ. A possible reason is that Ultron-PQ adds some new tokens to the vocabulary, thus general pre-training is necessary to understand the meaning of these tokens.

\begin{table}[!t]
    \centering
    \small
    \caption{Experiments about the memory cost and efficiency.}
    \begin{tabular}{l|p{0.18\linewidth}<{\centering}p{0.18\linewidth}<{\centering}p{0.18\linewidth}<{\centering}}
    \toprule
        Model & Memory & Latency & R@1 \\ 
    \midrule
        DocT5Query & 3.82MB & 5.63ms & 0.2327 \\
        DPR & 980MB & 18.87ms & 0.2908 \\
    \midrule
        Ultron-Atomic & 0 & 20.31ms & 0.3281 \\
        Ultron-URL & 51.0MB & 13.75ms & 0.2957 \\
        Ultron-PQ & 69.6MB & 8.90ms & 0.3155 \\
    \bottomrule
    \end{tabular}
    \label{tab:efficiency}
\end{table}
\subsection{Study of Memory and Efficiency (RQ3)}
Since document retrieval is a critical step in practical search applications, lower memory overhead and higher efficiency are necessary. We conduct experiments to compare the memory cost and inference latency of Ultron, DocT5Query, and DPR with brute-force search on MS MARCO. 

Observing the results in Table~\ref{tab:efficiency}, Ultron has a significant reduction of memory and inference latency compared to the DPR model, while achieving better results. Specifically, Ultron only needs to store the prefix tree, which spends 90\% less memory than the vectorized index of DPR. Most importantly, Ultron is more efficient than dual encoders as the latency drops from 18.87ms to 8.90ms with regard to 320K corpus size. Although the ANN approach can speed up dense retrieval, the latency of the dual encoder models will increase with the expansion of the corpus. But for Ultron, relevant docids are directly generated by the model through constrained beam search, where the speed is only related to the layer and width of the prefix tree.


\begin{figure}[!t]
    \centering
    \vspace{-0.1cm}
    \includegraphics[width=0.9\linewidth]{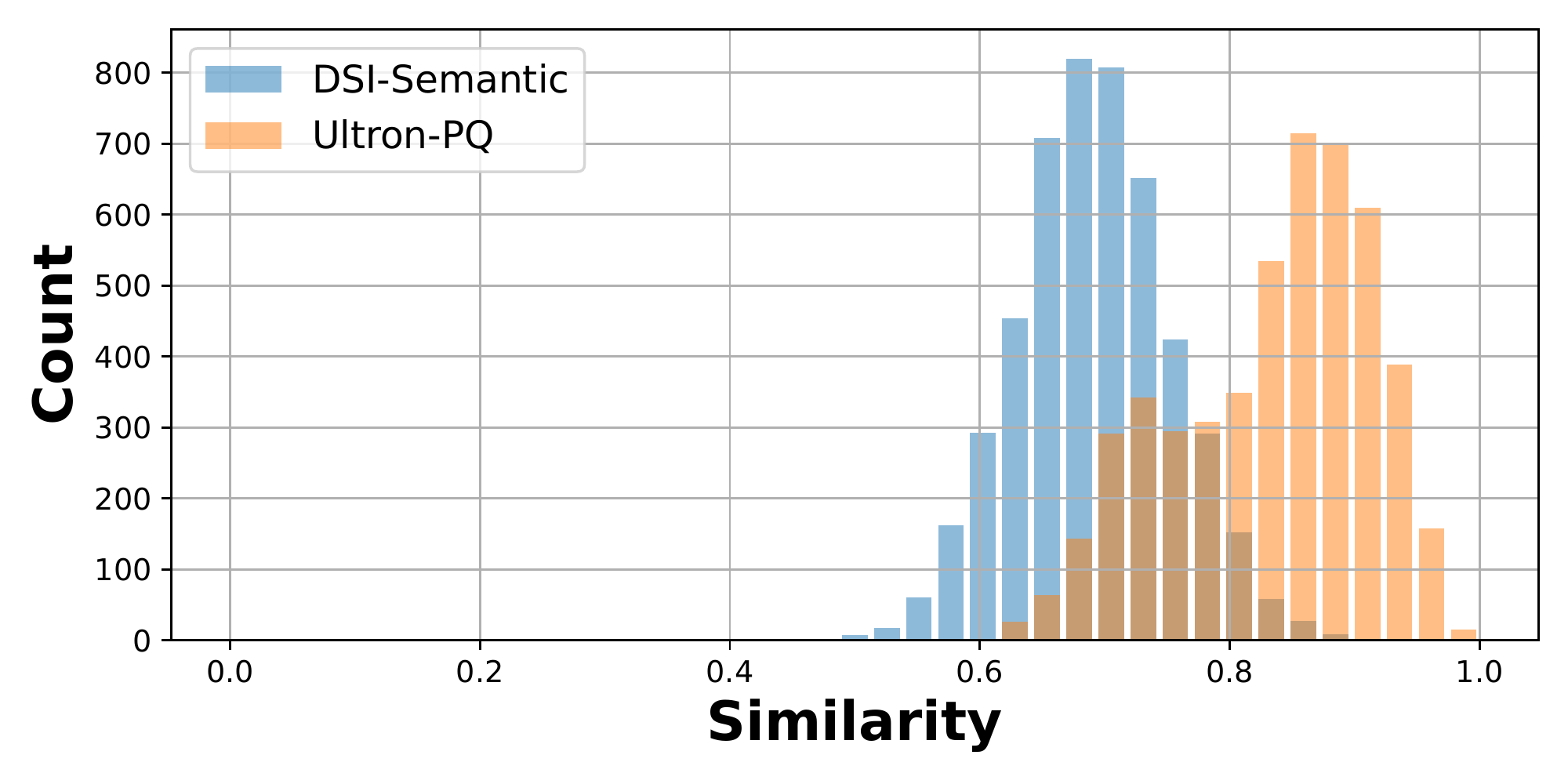}
    \caption{Distribution of similarity between documents.}
    \label{fig:docid_similarity}
\end{figure}
\subsection{Comparison of Different Docids (RQ4)}
Different docid representations will affect the inference difficulty of the end-to-end retriever. To observe the superiority of our designed semantic docids, we compare the docids used in DSI-Semantic and Ultron-PQ, which are both encoded based on pre-calculated document embeddings. Specifically, we randomly select 100 documents from the set of docids with the same prefixes (length $=2$), and calculate the similarity between the embeddings of every two documents. We divide the bar at 0.025 intervals, and count the number of document pairs that fall in each range.

From Figure~\ref{fig:docid_similarity}, it can be seen that the distribution of similarity between documents generally obeys normal distribution. The document embeddings of Ultron-PQ show higher consistency, which indicates that our semantics-based identifiers are more inclined to assign the same prefix to similar documents. We believe this is more conducive to the convergence of the model. Conversely, if similar documents are assigned different prefixes, the training and inference of the model will be more difficult and confusing.




\subsection{Limitation and Future Work}
Although we have achieved certain results in the model-centric paradigm, Ultron still faces several challenges to overcome. First, extending the model to web scale puts forward requirements for the design of the model-based indexer with higher capacity. Second, how to add new coming documents to the model-based indexer remains unexplored. On the basis of Ultron, there is plenty of room for future exploration. (1) We can combine the advantages of different docids or devise better docids to reinforce the model capacity for massive webpages. (2) A more reasonable model-based index structure, or the fusion of traditional and model-based indexes can be explored to deal with new documents.


\section{Conclusion}
In this work, we explore a novel model-centric paradigm for document retrieval. The model Ultron breaks away from the traditional index-based methods by encoding the knowledge of docids into an end-to-end model. Under the T5 backbone, we devise two types of semantic document identifiers, and a three-stage training strategy to optimize the model and adapt it to search scenarios. Experiments on two public datasets indicate the superiority of the model-based indexer on retrieval performance and efficiency over existing baselines.

\bibliography{refer}



\end{document}